\documentclass[useAMS,usenatbib]{mn2e}
\usepackage{times}
\usepackage{graphicx}
\usepackage{amssymb}
\usepackage{longtable}

\title[Why do galaxies stop forming stars?]{Why do galaxies stop forming stars? I. The passive fraction - black hole mass relation for central galaxies}
\author[Asa F. L. Bluck et al.]{Asa F. L. Bluck$^{1,*}$, Sara L. Ellison$^{1}$, David R. Patton$^{2}$, Luc Simard$^{3}$, J. Trevor Mendel$^{4}$, \newauthor Hossein Teimoorinia$^{1}$, Jorge Moreno$^{5,6}$, and Else Starkenburg$^{7,8}$
\\$^1$ Department of Physics and Astronomy, University of Victoria, 3800 Finnerty Road, Victoria, British Columbia, V8P 1A1, Canada
\\$^2$ Department of Physics and Astronomy, Trent University, 1600 West Bank Drive, Peterborough, Ontario, K9J 7B8, Canada 
\\$^3$ National Research Council of Canada, Herzberg Institute of Astrophysics, 5071 West Saanich Road, Victoria, British Columbia, V9E 2E7, Canada
\\$^4$ Max-Planck-Institut f\"ur Extraterrestrische Physik (MPE), Giessenbachstrasse, D-85748 Garching, Germany 
\\$^5$ Department of Physics and Astronomy, California State Polytechnic University Pomona, Pomona, CA 91768, USA
\\$^6$ TAPIR, Mailcode 350-17, California Institute of Technology, Pasadena, CA 91125, USA
\\$^7$ Schwarzschild Fellow, Leibniz-Institut f\"ur Astrophysik Potsdam (AIP), An der Sternwarte 16, 14482 Potsdam, Germany
\\$^8$ CIfAR Global Scholar
\\$*$ Email: abluck@uvic.ca }

\begin{document}

\maketitle

\begin{abstract}

\noindent We derive the dependence of the fraction of passive central galaxies on the mass of their supermassive black holes for a sample of over 400,000 SDSS galaxies at z $<$ 0.2. Our large sample contains galaxies in a wide range of environments, with stellar masses $8 < {\rm log}(M_{*}/M_{\odot}) < 12$, spanning the entire morphological spectrum from pure disks to spheroids. We derive estimates for the black hole masses from measured central velocity dispersions and bulge masses, using a variety of published empirical relationships. We find a very strong dependence of the passive fraction on black hole mass, which is largely unaffected by the details of the black hole mass estimate. 
Moreover, the passive fraction relationship with black hole mass remains strong and tight even at fixed values of galaxy stellar mass ($M_{*}$), dark matter halo mass ($M_{\rm halo}$), and bulge-to-total stellar mass ratio ($B/T$). Whereas, the passive fraction dependence on $M_{*}$, $M_{\rm halo}$ and $B/T$ is weak at fixed $M_{BH}$. These observations show that, for central galaxies, $M_{BH}$ is the strongest correlator with the passive fraction, consistent with quenching from AGN feedback.

\end{abstract}
\begin{keywords}
Galaxies: formation, evolution, AGN, morphology; black holes; star formation 
\end{keywords}

\section{Introduction}

Understanding why galaxies stop forming stars is an important unresolved question in galaxy evolution. Only $\sim$10\% of baryons reside within galaxies (e.g. Fukugita \& Peebles 2004, Shull et al. 2012), yet since galaxies lie at nodes in the cosmic web corresponding to local minima in the gravitational potential well, naively one would expect far more baryons to collate in galaxies, ultimately forming more stars. Theoretical models offer a wide range of solutions to this problem, relying on the physics of gas, stars, and black hole accretion disks as so called `baryonic feedback'. However, observational studies are required to test these models and provide evidence for their range and applicability. 

Since the discovery that most galaxies contain a supermassive black hole (e.g. Kormendy \& Richstone 1995), the energy released from forming these objects has become a popular mechanism for regulating gas flows and star formation in simulations, particularly for more massive galaxies (e.g. Croton et al. 2006, Bower et al. 2006, 2008, Somerville et al. 2008, Guo et al. 2011, Vogelsberger et al. 2014a,b, Schaye et al. 2015). In fact, substantial feedback from accretion around supermassive black holes is required in cosmological semi-analytic models, semi-empirical models, and hydrodynamical simulations to achieve the steep slope of the high-mass-end of the galaxy stellar mass function (e.g. Henriques et al. 2014, Schaye et al. 2015, Vogelsberger et al. 2014a,b). 

Observationally, predominantly only very indirect means for linking the star formation in large populations of galaxies to their AGN have been found, e.g. the strong dependence of the passive fraction on galaxy stellar mass (e.g. Baldry et al. 2006, Peng et al. 2010, 2012) and galaxy structure (e.g. Driver et al. 2006, Cameron et al. 2009, Cameron \& Driver 2009, Bluck et al. 2014). These galaxy properties both correlate (weakly) with black hole mass and, hence, the total energy released in forming the black hole (e.g. Silk \& Rees 1998, Fabian 1999, Bluck et al. 2011, 2014). More direct measurements of AGN driven winds in galaxies and radio bubbles in galaxy haloes have provided evidence for the mechanisms of AGN feedback, but typically only for a very small number of galaxies (e.g. McNamara et al. 2000, Nulsen et al. 2005, McNamara et al. 2007, Dunn et al. 2010, Fabian 2012, Cicone et al. 2013).

More recent work has linked the passive fraction of large populations of galaxies to the central density within 1 kpc (Cheung et al. 2012, Fang et al. 2013, Woo et al. 2014) and to the mass of the galactic bulge (Bluck et al. 2014, Lang et al. 2014). Both of these galaxy properties are expected to correlate strongly with the mass of the central black hole (e.g. Haring \& Rix 2004, McConnell \& Ma 2013) and hence may provide qualitative support for AGN feedback driven quenching. 

We find in Bluck et al. (2014) that `bulge mass is king' in the sense that bulge mass is a tighter and steeper correlator to the passive fraction for centrals than any other variable considered (including stellar mass, halo mass, disk mass, local density, and galaxy structure). We argue for an explanation of this observation through AGN feedback, relying on the tight relationships between the bulge and central black hole (e.g. Haring \& Rix 2004). The primary motivation for this Letter is to expand on that work by testing the role of AGN feedback in quenching central galaxies more directly. 

In this Letter we estimate black hole masses from their bulge masses and central velocity dispersions, using a wide range of published empirical relationships with dynamical measurements of black hole masses (e.g. McConnell \& Ma 2013). From this, we derive the relationship between passive fraction and estimated black hole mass. Our goal is to test whether AGN feedback is a viable route for quenching galaxies, and test whether or not it is dominant for a large sample of central galaxies. Throughout this Letter we assume a $\Lambda$CDM cosmology with H$_{0}$ = 70 km s$^{-1}$, $\Omega_{m}$ = 0.3, $\Omega_{\Lambda}$ = 0.7, and adopt AB magnitude units.

\section{Data Overview \& Passive Fraction}

We use the Sloan Digital Sky Survey Data Release 7 (SDSS DR7, Abazajian et al. 2009) spectroscopic sample as our data source. From this we collate a sample of 538046 galaxies (423480 centrals and 114566 satellites) with  $8 < {\rm log}(M_{*}/M_{\odot}) < 12$ at z $<$ 0.2. In this Letter (the first paper in the series) we concentrate on the central galaxies. The star formation rates (SFR) for these galaxies are calculated from emission lines for non-AGN star forming galaxies and from the strength of the 4000 \AA \hspace{0.05cm} break for non-emission line galaxies and AGN (Brinchmann et al. 2004). A fibre correction is applied based on galaxy colour and magnitude outside the aperture. The stellar masses for the galaxies, and their component disks and spheroids, are derived in Mendel et al. (2014), based on SED fitting to a dual S\'{e}rsic fit of the $ugriz$ wavebands (Simard et al. 2011). 
An n = 4 bulge and n = 1 disk model is used, and we test the reliability of this approach in Bluck et al. (2014) via model data. We define the galaxy structure (or morphology) to be $B/T = M_{\rm bulge} / M_{*}$, where $M_{*}$ is the total stellar mass of the galaxy.

The velocity dispersions are derived from the widths of absorption lines taken from Bernardi et al. (2003) with an updated method implemented as in Bernardi et al. (2007) to the later data releases. Velocity dispersions from absorption lines with a S/N $<$ 3.5 are discarded from our sample, and those with $\sigma$ $<$ 70 km/s are deselected for some analyses, due to the instrumental resolution of the SDSS spectra. We also restrict our final sample to galaxies with an error on the velocity dispersion of $\sigma_{\rm err}$ $<$ 50 km/s. We recover $\sim$ 80 \% of our parent sample which pass these data quality cuts. 

The halo masses used in this Letter are derived from an abundance matching technique applied to the total stellar mass of the group or cluster (from Yang et al. 2007, 2008, 2009). Testing of the group finding algorithm on model galaxies from the Millennium Simulation (Springel et al. 2005) showed that over 90 \% of galaxies are correctly assigned to groups at $M_{\rm halo} > 10^{12} M_{\odot}$. Using these group catalogues, centrals are defined as the most massive galaxy in the group, with satellites being any other galaxy within the group.

We follow the prescription for defining the passive fraction in Bluck et al. (2014). We define a galaxy to be passive if it is forming stars at a rate a factor of ten times lower than (emission line, non-AGN) star forming galaxies matched at the same stellar mass and redshift. This cleanly divides the two peaks of the bimodal distribution in SFR at all mass ranges. The passive fraction is the ratio of passive-to-total galaxies in each binning. We correct for the flux limit of the SDSS by weighting each galaxy in the passive fraction by the inverse of the volume over which its $ugriz$ magnitudes would pass all of the selection criteria. The errors on the passive fraction are computed in this work via the jack-knife technique. See Bluck et al. (2014) \S 2 \& 3 for full details on these data and techniques.

\section{Black Hole Masses}

Today there are less than 100 reliable dynamical measurements of central black hole masses in existence (see McConnell \& Ma 2013 for a compilation). This is far too few to investigate the dependence of the passive fraction on black hole mass directly, due in part to the statistical nature of the passive fraction and the need for a wide range in galaxy properties. Fortunately, there exist tight relationships between dynamical measurements of central black hole masses and the properties of their host galaxies, particularly the bulge or spheroid in which the black hole resides (e.g. Magorrian et al. 1998; Ferrarese \& Merritt 2000; Gebhardt et al. 2000; Haring \& Rix 2004). On a case by case basis, using these relationships to estimate a given galaxy's black hole mass would lead to a high level of uncertainty, but for very large samples the average values should yield reliable measurements, in the regime where the relationships are tested.

In this work, we use a number of the galaxy - black hole relationships from the literature to estimate black hole masses. Specifically, we use central velocity dispersion ($\sigma_{c}$), stellar mass of the bulge or spheroid ($M_{\rm bulge}$), and combinations of these two. We also test separating the sample into early- and late-type systems and using separate scaling laws for each (as suggested in McConnell et al. 2011). The formulae used to estimate the black hole masses, and the references for these, are all presented in Table 1. 

For the velocity dispersions, we first make an aperture correction, so that all measurements are made at the same effective aperture. We use the formula in Jorgensen et al. (1995), specifically calculating:

\begin{equation}
\sigma_{c} = \big(\frac{R_{c}}{R_{\rm app}}\big)^{-0.04} \sigma_{\rm app}
\end{equation}

\noindent where the central radius, $R_{c}$, is chosen to be in line with the measurements made in the literature, with typical values of $R_{e}$ or $R_{e}/8$ used. $R_{e}$ is the bulge (or spheroid) effective radius, calculated in our sample through S\'{e}rsic index fitting (see Simard et al. 2011 for full details). We note that the aperture correction only affects the final mass estimate by typically $<$10\%. For the bulge masses no such correction is necessary, and thus, this provides a natural test to the reliability of the method.

Since our goal is to present the passive fraction dependence on black hole mass for all galaxies, we do not restrict the sample to pure spheroids. Thus, there is a substantial population of disks present in our sample. This poses no significant difficulty for measuring the black hole masses via bulge mass (but see Bluck et al. 2014 appendices for a discussion on their reliability). However, for the velocity dispersions, contamination in their measurement from rotating disks could be an issue. We test this in the next section by looking at early- and late-type galaxies separately, and comparing estimates of $M_{BH}$ from velocity dispersion methods to bulge mass methods. We additionally test the impact of the choice of black hole mass calibration by comparing a number of different techniques, noting a high level of consistency between them (see next section).

\begin{table*}
\caption{Black Hole Mass Estimates}
 \label{tab1}
 \begin{tabular}{@{}cccccc}
  \hline
\hline
Source & log($M_{BH}/M_{\odot}$) =  & Morphology & Dispersion (dex) & $N_{\rm Gal}$ & Fig. 1 Colour \\ 
\hline
Hopkins et al. (2007) & 0.54 $\times$ log($M_{\rm bulge} / 10^{11} M_{\odot}$) + 2.18 $\times$ log($\sigma_{c}$ / 200 km/s) + 8.24 & No Cut & 0.22 & 38 & Blue \\
Ferrarese \& Merritt (2000) & 4.80 $\times$ log($\sigma_{c}$) - 2.90 & No Cut & 0.31 & 12 & Green\\
Haring \& Rix (2004) & 1.12 $\times$ log($M_{\rm bulge} / 10^{11} M_{\odot}$) + 8.2 & No Cut & 0.33 & 30 & Cyan\\
McConnell \& Ma (2013) & 5.64 $\times$ log($\sigma_{c}$ / 200 km/s) + 8.32 & No Cut & 0.38 & 72 & Orange\\
McConnell \& Ma (2013) & 5.20 $\times$ log($\sigma_{c}$ / 200 km/s) + 8.39 & ETGs & 0.34 & 53 & Red$^{*}$\\
McConnell \& Ma (2013) & 5.06 $\times$ log($\sigma_{c}$ / 200 km/s) + 8.07 & LTGs & 0.46 & 19 & Red$^{*}$\\
\hline
\end{tabular}
$^{*}$ for the ETG/ LTG samples, galaxies defined as early types ($B/T >$ 0.5) are fit with the ETG formula and those defined as late types ($B/T <$ 0.5) are fit with the LTG formula, they are then recombined into the same sample, and shown in red in Fig. 1. $N_{\rm Gal}$ is the number of galaxies used in the dynamical fit.
\end{table*}

\begin{figure*}
\rotatebox{270}{\includegraphics[height=0.33\textwidth]{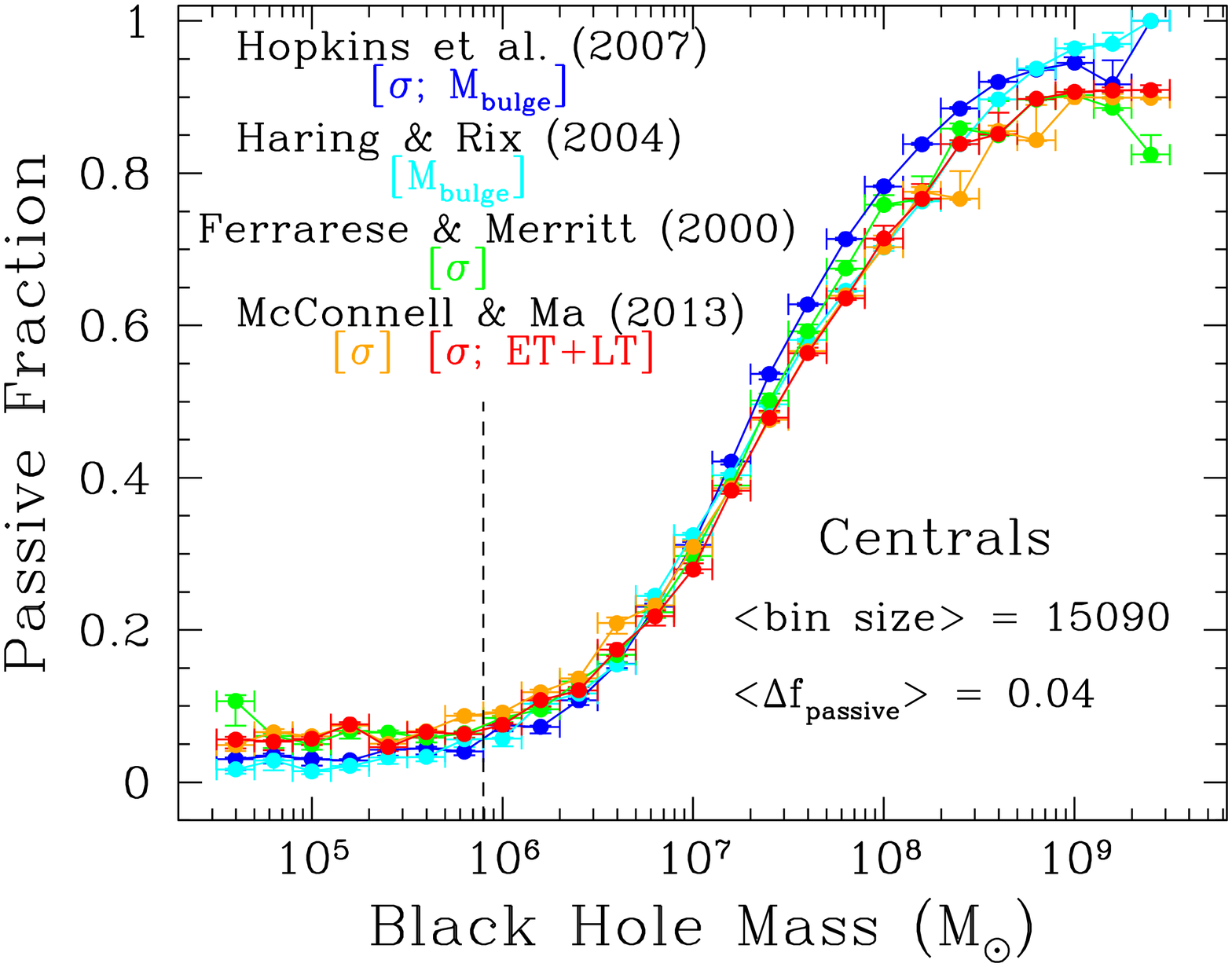}}
\rotatebox{270}{\includegraphics[height=0.33\textwidth]{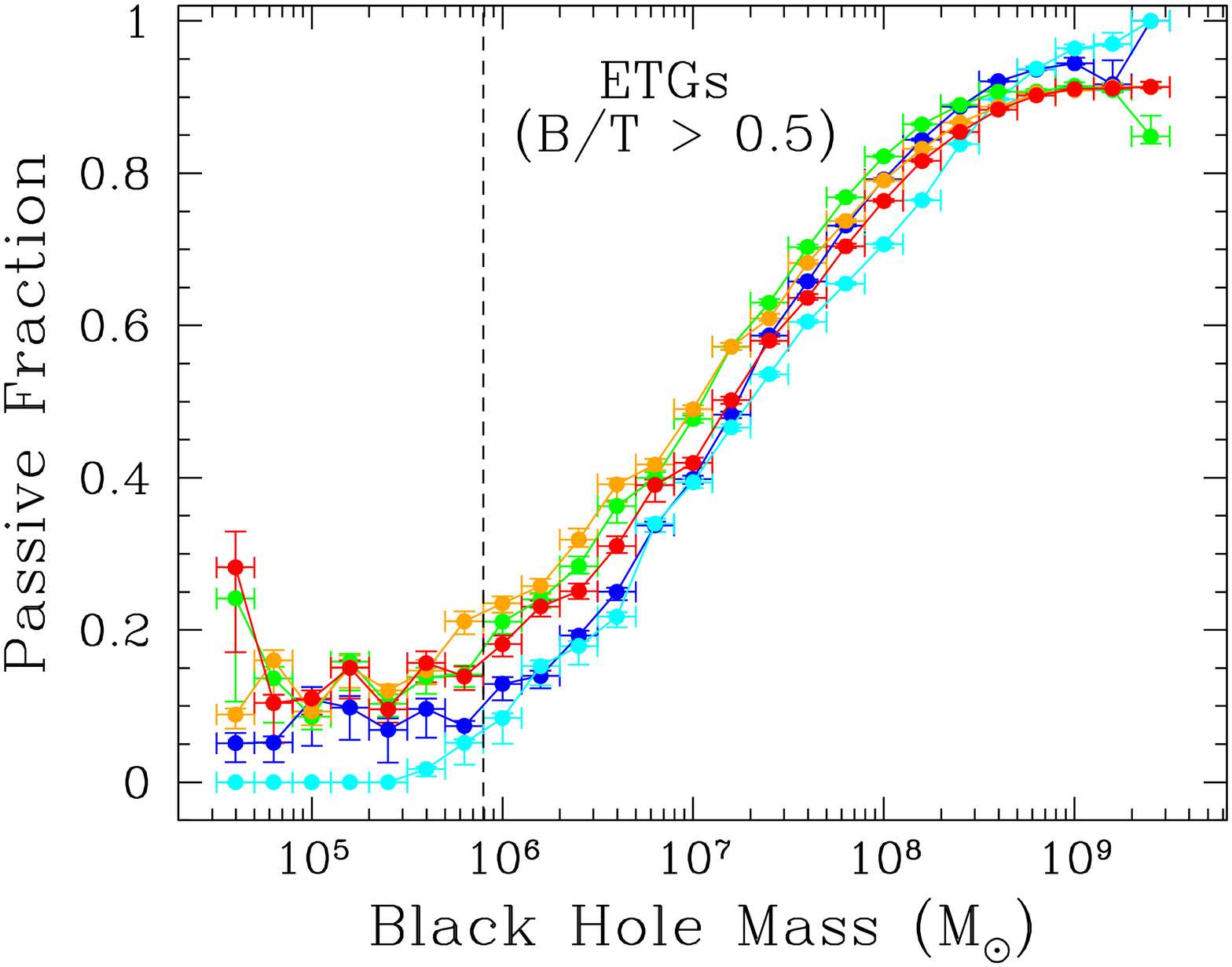}}
\rotatebox{270}{\includegraphics[height=0.33\textwidth]{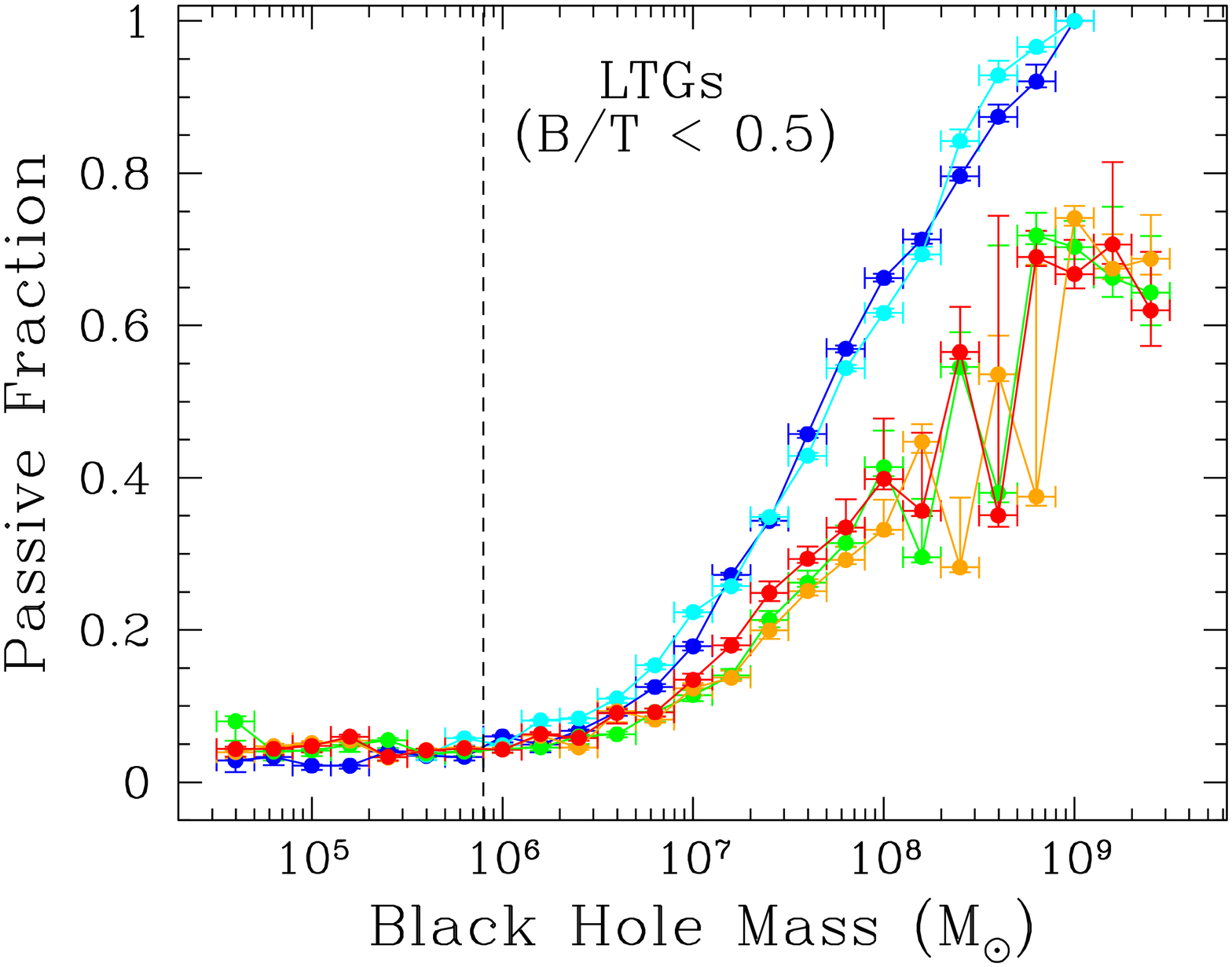}}
\caption{The passive fraction - black hole mass relationship for all central galaxies (left), early-type central galaxies (middle), and late-type central galaxies (right). The black hole masses are derived from a variety of different techniques, including central velocity dispersion and bulge mass methods (see Table 1 for details). It is clear that there is a steep dependence of the passive fraction on black hole mass for centrals, regardless of morphology. There is generally very good agreement in the passive fraction - black hole mass relationship between the different methods used for computing the black hole masses for both the full morphological sample (left) and early-type galaxies (middle). 
However, for late-type galaxies (right) there is disagreement between the methods that incorporate bulge mass (blue lines) and those that use velocity dispersion alone (other colours). We comment on this in \S 4.1.  The average difference in the passive fraction between the black hole mass estimates is shown for the full sample on the left panel. The error bars shown on all plots represent the bin size and 1 $\sigma$ error on the passive fraction computed via the jack-knife technique. }
\end{figure*}

\section{Results}

\subsection{The $f_{\rm passive} - M_{BH}$ Relationship for Central Galaxies}

In this section we compute, for the first time, the relationship between the fraction of passive galaxies and the mass of their central supermassive black holes, for a population of over 400,000 central galaxies (see Fig. 1). The left panel shows the $f_{\rm passive} - M_{BH}$ relationship for all central galaxies, which are defined as the most massive group members in the catalogues of Yang et al. (2009). We show this relationship via a number of different black hole mass calibrations, summarized in Table 1. These include relationships from various authors based on central velocity dispersion, bulge mass, and a combination of the two. For the full morphological sample, there is remarkably good agreement between all of the methods used for estimating black hole masses.

The $f_{\rm passive} - M_{BH}$ relationship is very steep for central galaxies, with galaxies with low black hole masses being predominantly star forming and galaxies with high black hole masses being predominantly passive. The cross-over mass, where 50\% of galaxies are passive occurs at $M_{BH} \sim 10^{7.5} M_{\odot}$. Values of black hole mass $< 10^{6} M_{\odot}$ are highly uncertain due to the $M_{BH} - \sigma$ , $M_{\rm bulge}$ relations not extending into this domain. Additionally, velocity dispersions $<$ 70 km/s are less reliable than for the rest of the sample (e.g. Bernardi et al. 2007), and this corresponds to approximately $M_{BH}$ $< 10^{6} M_{\odot}$ as well. For these reasons, we advise readers to treat the low $M_{BH}$ values more as upper limits (this threshold is indicated with a dashed vertical line in all relevant plots).

In the middle and right panels of Fig. 1, we show the passive fraction - black hole mass relationship for centrals split into early-types ($B/T >$ 0.5) and late-types ($B/T <$ 0.5) respectively (computed in Mendel et al. 2014). For the early-type sample, all of the black hole mass estimates lead to very similar passive fraction relationships, as with the full sample. However, for the late-type galaxies, there is some disagreement between methods which use velocity dispersion alone and those which incorporate bulge mass. Specifically, the passive fraction is lower for velocity dispersion estimates of black hole mass than with bulge mass estimates. 
This is most probably explained by the velocity dispersions being contaminated by disk rotation into the plane of the sky, and by measurements of the absorption line widths being contaminated by multiple emission lines obscuring the true continua. Both of these effects can lead to an over-estimate of the black hole mass, and hence lead to lower passive fractions for the measured black hole masses from velocity dispersion methods compared to bulge mass methods. 

For bulge mass only methods contamination of the kinematic measurement in disks is clearly not an issue, but interestingly, in the case where we use $\sigma_{c}$ and $M_{\rm bulge}$ together (Hopkins et al. 2007), the contamination is also negligible (compare blue and cyan lines in Fig. 1, right panel). Thus, we posit that to estimate black hole masses for the full morphological sample (i.e. including disk galaxies) a measure of the mass of the bulge is {\it essential} to include. For the remainder of the analyses in this Letter we focus on the black hole mass estimates from bulge mass and velocity dispersion together (from Hopkins et al. 2007, see top row of Table 1), where kinematic contamination is not a significant issue. This technique draws on multiple complementary measurements of the central regions of galaxies, from spectroscopic kinematics and photometric profile and SED fitting. We note that all of our conclusions remain unchanged if we restrict to early-type galaxies only and use velocity 
dispersion methods to compute our black hole masses. The advantage of incorporating the bulge mass is to leave the sample complete.

In an upcoming paper in the series (Paper II, Bluck et al. in prep.) we show the passive fraction relationship with black hole mass for satellites. We find that, whereas centrals with low black hole masses are essentially all star forming, satellites with low black hole masses are more frequently passive. This hints at the need for additional quenching mechanisms to AGN feedback for satellites, but not necessarily for centrals.

\subsection{Is the $f_{\rm passive} - M_{BH}$ Relationship Fundamental for Central Galaxies?}

The steep dependence of the passive fraction on supermassive black hole mass for centrals (Fig. 1, left panel) is consistent with the quenching of central galaxies being driven by AGN feedback. However, given that there are observationally well known, and theoretically expected, correlations between the black hole mass (derived from $M_{\rm bulge}$ and/or $\sigma_{c}$) and other galaxy properties (e.g. $M_{\rm halo}$, $M_{*}$, $B/T$), it is not immediately clear whether the $f_{\rm passive} - M_{BH}$ relationship is fundamental or not (i.e. whether it is derivative of some other, more fundamental, parameter).

To test the possibility that another galaxy property is more fundamental to passivity than black hole mass, we look at the $f_{\rm passive} - M_{BH}$ relationship at a fixed values of the following galaxies properties: $M_{\rm halo}$, $M_{*}$, and $B/T$ (presented in Fig. 2, top row). We find that the $f_{\rm passive} - M_{BH}$ relationship remains tight and steep even at a fixed halo mass, stellar mass and galaxy structure ($B/T$). Out of these three variables, we find the greatest variation at a fixed $M_{BH}$ with $B/T$, suggesting that the structure of a galaxy is an important secondary consideration for whether a galaxy will be forming stars or not. In Fig. 2 (bottom row) we present the $f_{\rm passive} - M_{\rm halo}, M_{*}$, and $B/T$ relationships, each at fixed $M_{BH}$. These are all remarkably flat compared to the $f_{\rm passive} - M_{BH}$ relationship, suggesting that it must be more fundamental than these alternatives.

Particularly, the $f_{\rm passive} - M_{\rm halo}$ relationship, at a fixed $M_{BH}$, is almost entirely flat, indicating essentially no dependence of galaxy quenching on the mass of the halo, once the correlation with black hole mass is accounted for. This has a profound implication for models of galaxy quenching. Methods that derive the energy to do work on the gas in a galaxy from the dark matter halo cannot be driving the quenching of centrals (e.g. halo mass quenching, Dekel \& Birnboim 2006, Dekel et al. 2009, Woo et al. 2013). However, at high $M_{BH}$ values, there is some hint of a subtle positive correlation of the passive fraction with $M_{\rm halo}$, which may be explained if AGN feedback is more efficient in higher mass haloes (e.g. Dekel et al. 2014).

The $f_{\rm passive} - M_{*}$ relationship at fixed $M_{BH}$ is either flat, or in certain $M_{BH}$ ranges, {\it negative}. This striking result clearly indicates that stellar mass is not the primary driver to passivity in central galaxies (as advocated in Baldry et al. 2006, Peng et al. 2010, 2012). This result further suggests that the quenching of central galaxies cannot be driven by stellar or supernova feedback, which would both correlate with $M_{*}$ (as integrated star formation rate) more strongly than with $M_{BH}$. The negative trend is most probably explained by the fact that increasing $M_{*}$ at fixed $M_{BH}$ results in an increase in $M_{\rm disk}$, and hence a decrease in $B/T$. This follows directly from the $M_{BH} - M_{\rm bulge}$ relationship, and we have seen already that $B/T$ is an important {\it secondary} correlator to the passive fraction.

The $f_{\rm passive} - M_{BH}$ relationship is also steeper and tighter than the $f_{\rm passive} - B/T$ relationship (which was suggested to be primary in Driver et al. 2006, Cameron et al. 2009). This implies that major galaxy mergers (which can destroy in situ disks, e.g., Cole et al. 2000) are not necessary for quenching central galaxies, but substantial central black holes are. However, see Hopkins et al. (2013) for evidence that disks can regrow in the merger process complicating this issue. Additionally, morphological quenching (e.g. Martig et al. 2009) cannot be the primary driver to the quenching of star formation in centrals, since black hole mass correlates much stronger with the passive fraction than galaxy structure. See Bluck et al. (2014) \S 5 for a more thorough discussion on the role of various processes in quenching central galaxies.

We consider the relative role of the supermassive black hole in quenching central and satellites galaxies compared to many other galaxy properties (including bulge mass, disk mass, and local density) in an upcoming work in this series (Paper III, Teimoorinia et al., in prep.). Black hole mass remains the most important single correlator to the passive fraction in this larger analysis as well.

\begin{figure*}
\rotatebox{270}{\includegraphics[height=0.33\textwidth]{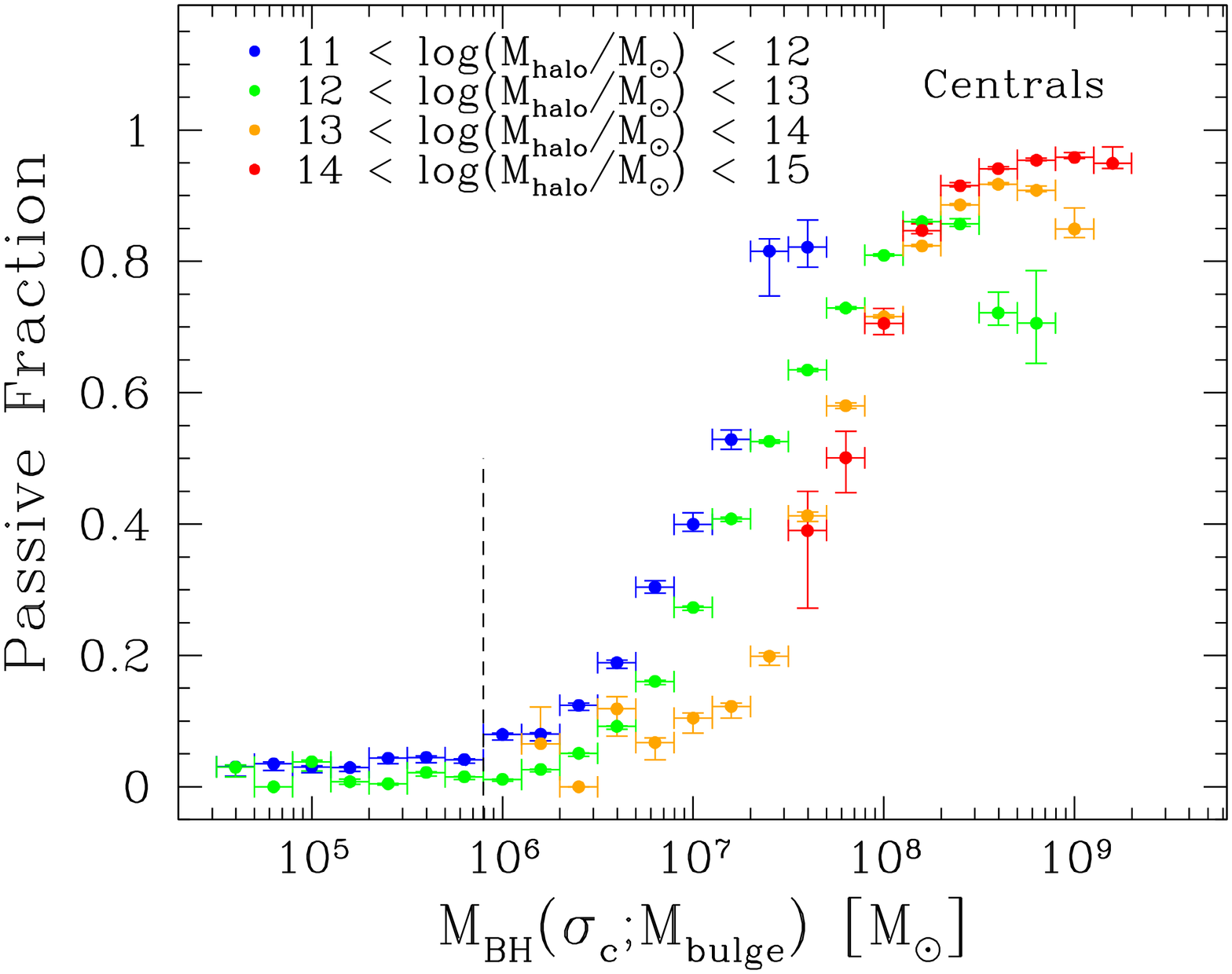}}
\rotatebox{270}{\includegraphics[height=0.33\textwidth]{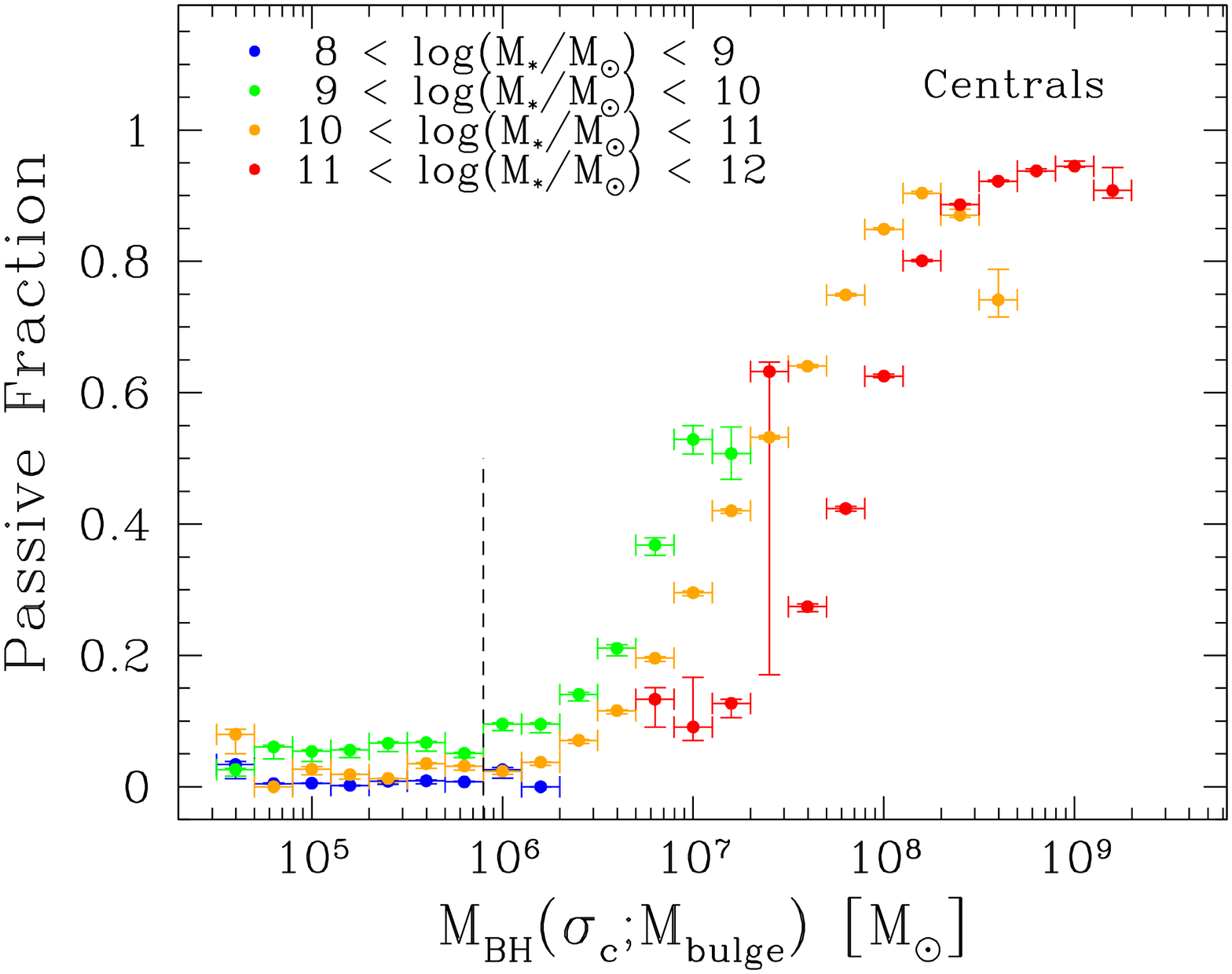}}
\rotatebox{270}{\includegraphics[height=0.33\textwidth]{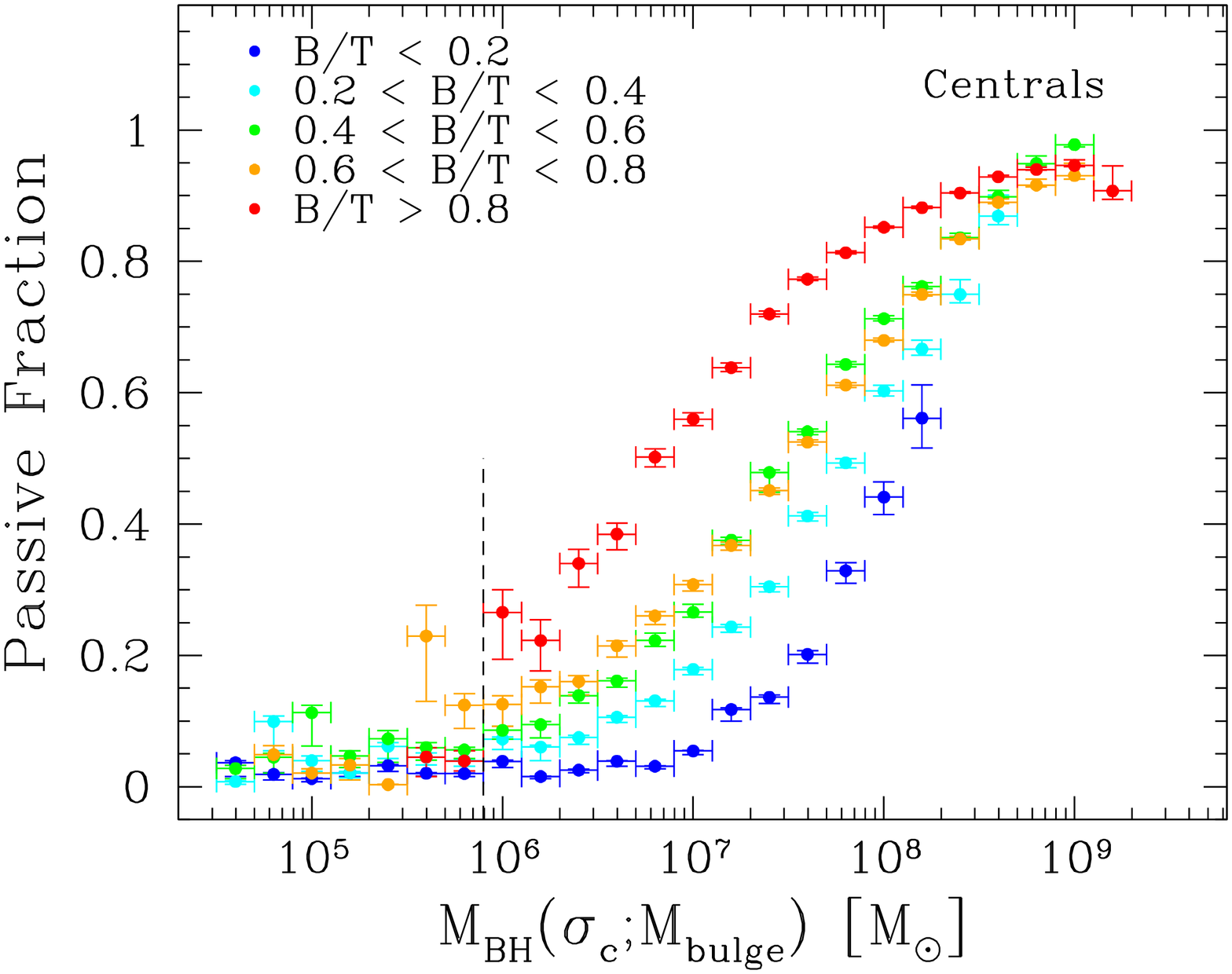}}
\rotatebox{270}{\includegraphics[height=0.33\textwidth]{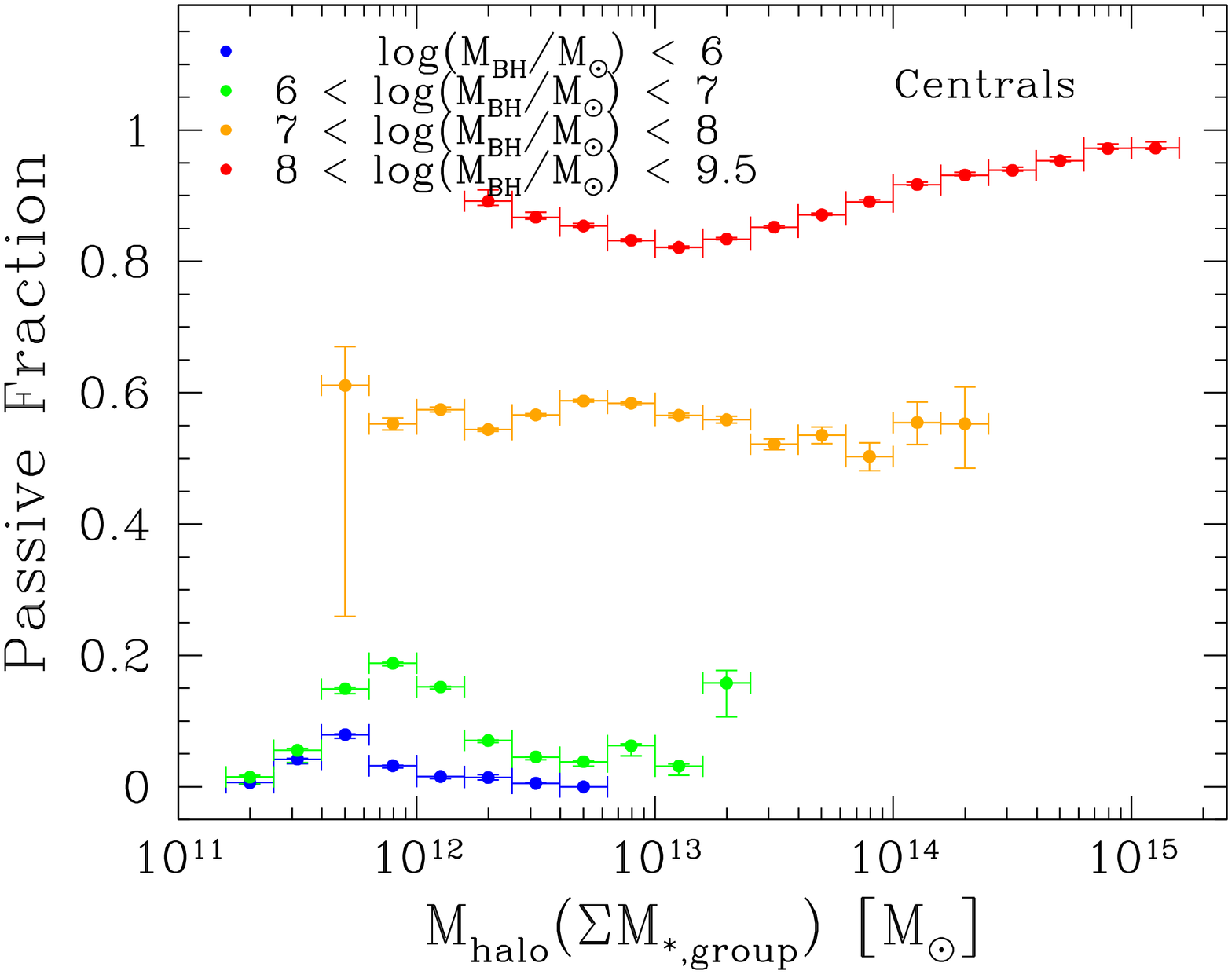}}
\rotatebox{270}{\includegraphics[height=0.33\textwidth]{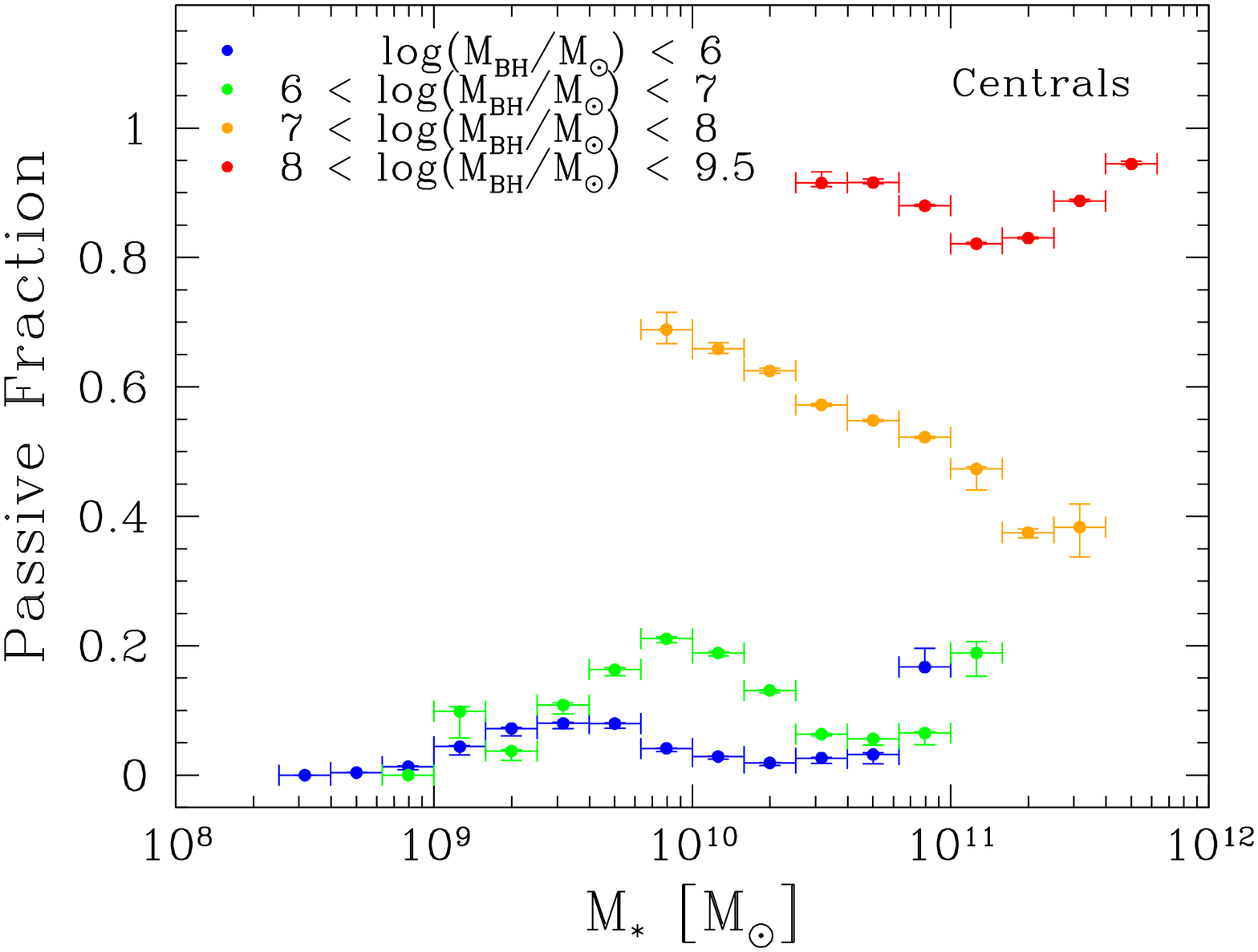}}
\rotatebox{270}{\includegraphics[height=0.33\textwidth]{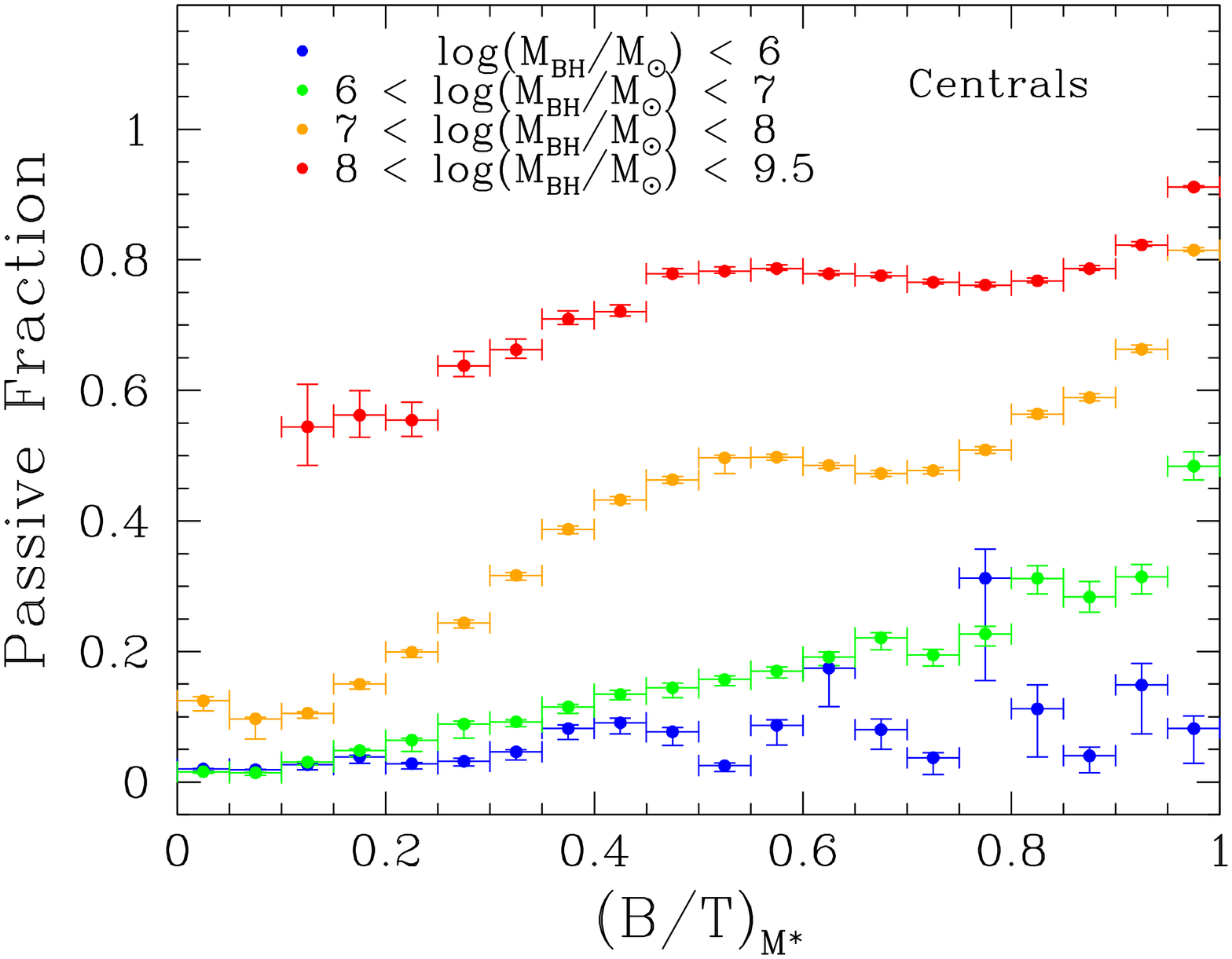}}
\caption{The passive fraction dependence for central galaxies on: {\it top row} - black hole mass, subdivided (from left to right) by group halo mass, total stellar mass, and bulge-to-total stellar mass ratio ($B/T$); {\it bottom row} - group halo mass, total stellar mass, and $B/T$, each subdivided by black hole mass. The black hole masses are estimated from measured bulge masses and central velocity dispersions, according to the empirical relationship of Hopkins et al. (2007), see \S 3 and Table 1. Black hole masses to the left of the vertical dashed lines (top row) should be treated as upper limits. The halo masses are inferred from an abundance matching technique applied to the total stellar mass of group (Yang et al. 2009), see \S 2. By comparing each column, we directly see that black hole mass is a much steeper and tighter correlator to the passive fraction than halo mass, stellar mass, or galactic structure ($B/T$). 
Thus, the quenching of central galaxies is intimately connected to the mass of their central black holes, most probably implicating AGN feedback as the dominant quenching mechanism. The error bars shown represent the bin size and 1 $\sigma$ error on the passive fraction computed via the jack-knife technique. }
\end{figure*}

\section{Conclusions}

In this Letter, we present for the first time the relationship between the fraction of passive galaxies and the mass of their supermassive black holes, for a sample of over 400,000 central galaxies at z $<$ 0.2 (Fig. 1). We find that the $f_{\rm passive} - M_{BH}$ relationship is very steep, with the vast majority of galaxies with low black hole masses being star forming and the vast majority of galaxies with high black hole masses being passive. We find a cross-over mass (where 50\% of galaxies are passive) at $M_{BH} \sim 10^{7.5} M_{\odot}$. The steep dependence of the passive fraction on black hole mass is consistent with AGN feedback being the dominant mechanism for quenching central galaxies. We test splitting our sample into early- and late-types, finding generally very good accord between all of the black hole mass estimates, except for late-type galaxies. Here we find that incorporating bulge mass into the black hole mass estimate is essential because velocity dispersion alone methods are vulnerable 
to contamination from disk rotation.

Finally, we find that black hole mass is a more fundamental correlator to the passive fraction than total stellar mass, group halo mass, or galaxy structure (Fig. 2). This is highly suggestive of the quenching of central galaxies being driven by AGN feedback. However, it is not inconceivable that the mechanism which grows the central region of galaxies (and presumably their supermassive black holes as well) is directly responsible for the quenching of central galaxies. Yet this would have to occur in such a way as to leave the central density as the primary correlator to the passive fraction, with $M_{\rm halo}$, $M_{*}$, and $B/T$ being substantially less important. We conclude that this is less likely than AGN feedback being responsible.


\end{document}